\documentclass[twocolumn,twoside,slac_two]{revtex4}
\pdfoutput=1
\usepackage{graphicx}
\usepackage{bm}
\usepackage{fancyhdr}
\usepackage{natbib}
\begin{document}

%Title of paper
\title{High-energy atmospheric neutrinos}

\author{S. I. Sinegovsky}
\affiliation{Irkutsk State University, Irkutsk, 664003, Russia}
\author{A. A. Kochanov}
\affiliation{Institute of Solar-Terrestrial Physics, Russian Academy of Sciences, Irkutsk, 664033, Russia}

\author{T. S. Sinegovskaya}
\affiliation{Irkutsk State Railway University, Irkutsk, 664046, Russia}

\begin{abstract}
        
High-energy neutrinos, arising from decays of mesons that were produced through the cosmic rays collisions with air nuclei, form unavoidable background noise in the astrophysical neutrino detection problem. The atmospheric neutrino flux  above 1 PeV should be supposedly dominated by the contribution of charmed particle decays.  These (prompt) neutrinos originated from decays of massive and shortlived particles,  $D^\pm$, $D^0$, $\overline{D}{}^0$, $D_s^\pm$, $\Lambda^+_c$, form the most uncertain fraction of the high-energy atmospheric  neutrino flux because of poor explored processes of the charm production.  Besides, an ambiguity in high-energy behavior of  pion and especially kaon production cross sections for nucleon-nucleus collisions may affect essentially the calculated neutrino flux. There is the energy region where above flux uncertainties superimpose.

A new calculation presented here reveals sizable differences, up to the factor of 1.8 above  1 TeV, in  muon neutrino flux predictions obtained with usage of known hadronic models, SIBYLL 2.1 and QGSJET-II. The atmospheric neutrino flux in the energy range $10-10^7$ GeV is computed within 1D approach to solve nuclear cascade equations in the atmosphere, which takes into account non-scaling behavior of the inclusive cross-sections for  the particle production, the rise of total inelastic hadron-nucleus cross-sections and nonpower-law character of the primary cosmic ray spectrum. This approach was recently tested in the atmospheric muon flux calculations \cite{kss08}.  The results of the neutrino flux calculations are compared with the Frejus, AMANDA-II and IceCube measurement data.   
\end{abstract}

\maketitle

\thispagestyle{fancy}

\section{Introduction}

Atmospheric neutrinos (AN) appear in decays of mesons (charged pions, kaons etc.) produced through collisions of  high-energy cosmic rays with air nuclei.  The AN flux in the wide energy range remains the issue of the great interest  since the low energy AN flux is a research matter in the neutrino  oscillations studies, and the high energy atmospheric neutrino flux is now appearing as the unavoidable background for astrophysical neutrino experiments~\cite{nt200-06, nt200-09, amanda07, amanda08, icecube, amanda09,  amanda10,  antares09}.  To present day a lot of AN flux calculations are made, among which~\cite{Volk80, Dedenko89, Lipari93,  nss98, FNV2001,  BGLRS04, HKKM04, DM, kss09} (see  also~\cite{Naumov2001, GH} for a review of 1D and 3D calculations of the  AN flux), but so far we don't know,  how discrepancy is strong in the conventional  neutrino flux  resulted from various hadronic interaction models, how  much  differences are great due to uncertainties in primary cosmic ray spectra and composition in the ``knee'' region.

In this work we present results of new one-dimensional calculation of the atmospheric muon neutrino flux in the range $10$--$10^7$ GeV made with use of the hadronic models  QGSJET-II 03~\cite{qgsjet2}, SIBYLL 2.1~\cite{sibyll} as well as the model by Kimel \&  Mokhov (KM)~\cite{KMN} that were tested also in recent atmospheric muon flux calculations~\cite{kss08,sks10}. We  compute here zenith-angle distribution of the conventional neutrinos and compare calculated neutrino energy spectra with the data of AMANDA-II and IceCube experiments.

\section{The method and input data \label{sec:method}} 

The calculation is performed on the basis of the  method~\cite{NS} of solution of the hadronic cascade equations in the atmosphere,  which takes into account non-scaling behavior of inclusive particle production cross-sections, the rise of total inelastic hadron-nuclei cross-sections, and the non-power law primary spectrum (see also~\cite{kss08, kss09, sks10}).  As the primary cosmic ray spectra and composition in wide energy range used is the model recently proposed  by Zatsepin \& Sokolskaya (ZS)~\cite{ZS3C}, which fits well the  ATIC-2 experiment data~\cite{atic2} and supposedly to be valid up to $100$ PeV. The ZS proton spectrum at $E\gtrsim 10^6$ GeV is compatible with KASCADE data~\cite{KASCADE05} as well the helium  one is  within the range of the KASCADE spectrum obtained with the usage of QGSJET 01 and SIBYLL models. Alternatively in the energy  range $1-10^6$ GeV we use the parameterization by Gaisser, Honda, Lipari and Stanev (GH)~\cite{GH}, the version with the high fit to the helium data. Note this version is consistent with the data of the  KASCADE experiment at $E_0>10^6$ GeV that was obtained (through the EAS simulations) with the SIBYLL 2.1.
To illustrate the distinction of the hadron models employed in the computations, it is appropriate to compare the spectrum-weighted moments (Table~\ref{tab_z}) computed for proton-air interactions (for $\gamma =1.7$): 
 \begin{equation}\label{eq:mom}
z_{pc}(E_0)%= \int\limits_0^1 x^{\gamma-1}\frac x{\sigma^{in}_{pA}}\frac {d\sigma_{pc}}{dx}dx
=\int\limits_0^1\frac {x^{\gamma}}{\sigma^{in}_{pA}}\frac {d\sigma_{pc}}{dx}\,dx,
\end{equation}
where  $x=E_c/E_0$, \ $c=p,n,\pi^\pm, K^\pm$.
The values in Table~\ref{tab_z} display approximate scaling law  both  in SIBYLL 2.1 and KM  and little violation of the scaling  in the QGSJET-II for $p$ and $\pi^\pm$.	

\begin{table}[t]
%\begin{center}
\caption{Spectrum weighted moments $z_{pc}(E_0)$ calculated for $\gamma=1.7$}
%  \centering % \vspace{0.2cm} 
  \begin{tabular}{|c|c|c|c|c|c|c|c|}
  \hline 
   Model  & $E_0$, GeV & $z_{pp}$ & $z_{pn}$ & $z_{p\pi^+}$  & $z_{p\pi^-}$  & $z_{pK^+}$  & $z_{pK^-}$  \\
   \hline 
             & $10^2$     & 0.174    & 0.088    & 0.043     & 0.035  & 0.0036   &  0.0030   \\
 QGSJET   & $10^3$     & 0.198    & 0.094    & 0.036     & 0.029  & 0.0036   &  0.0028   \\
  II-03   & $10^4$     & 0.205    & 0.090    & 0.033     & 0.028  & 0.0034   &  0.0027   \\
            \hline
             & $10^2$     & 0.211    & 0.059    & 0.036     & 0.026  & 0.0134   &  0.0014   \\
 SIBYLL   & $10^3$     & 0.209    & 0.045    & 0.038     & 0.029  & 0.0120   &  0.0022   \\ 
  2.1     & $10^4$     & 0.203    & 0.043    & 0.037     & 0.029  & 0.0097   &  0.0026   \\ 
            \hline  
          & $10^2$     & 0.178    & 0.060    & 0.044     & 0.027  & 0.0051  &  0.0015   \\
   KM     & $10^3$     & 0.190    & 0.060    & 0.046     & 0.028  & 0.0052  &  0.0015   \\ 
          & $10^4$     & 0.182    & 0.052    & 0.046     & 0.029  & 0.0052  &  0.0015   \\   
            \hline%\hline                         
  \end{tabular}
\label{tab_z}
%\end{center} 
  \end{table} 
%====================================================

\section{Atmospheric muon neutrino flux \label{sec:an}}

Along with major sources of the muon neutrinos, $\pi_{\mu2}$ and  $K_{\mu2}$ decays, we consider three-particle semileptonic decays, $K^{\pm}_{\mu3}$, $K^{0}_{\mu3}$,  the contribution originated from decay chains   $K\rightarrow\pi\rightarrow\nu_\mu$ ($K^0_S\rightarrow \pi^+\pi^-$, $K^\pm \rightarrow \pi^\pm \pi ^0$), as well as small fraction from the muon decays. % $\mu_{e3}$ .

One can neglect the 3D effects in calculations of the atmospheric muon neutrino flux  
near vertical at energies $E \gtrsim 1$ GeV and  at $E \gtrsim 5$ GeV in case of directions close to horizontal (see~\cite{BGLRS04,HKKM04}).

A comparison of  ($\nu_\mu+\bar\nu_\mu$) flux calculations for the three hadronic models under study is made in 
Table~\ref{tab_flux}: column 1, 2 and 3  presents the flux ratio, $\phi_{\nu_{\mu}}^{(\rm       SIBYLL)}/\phi_{\nu_\mu}^{(\rm KM)}$,  $\phi_{\nu_{\mu}}^{(\rm QGSJET{\text -}II)}/\phi_{\nu_\mu}^{(\rm KM)}$ and  $\phi_{\nu_{\mu}}^{(\rm SIBYLL)}/\phi_{\nu_{\mu}}^{\rm(QGSJET{\text -}II)}$ correspondingly, calculated at $\theta=0^\circ$ and $90^\circ$ (in brackets) with usage of the GH and ZS primary spectrum.  
  \begin{table}[!ht]
  \caption{Ratio of the $\nu_\mu$ fluxes at $\theta=0^\circ$\,($90^\circ$) calculated with the
    SIBYLL 2.1, QGSJET-II, and KM}
  \label{tab_flux}
%  \centering\vspace{0.2cm} 
  \begin{tabular}{|c|c|c|c|} \hline
     $E_{\nu}$, GeV  & $1$ & $ 2 $ & $3$ \\ 
   \hline %\hline 
    & \multicolumn{3}{|c|}{GH} \\ 
    $10^2$  & 1.65 (1.22)  & 0.97 (0.85)  & 1.65 (1.36) \\
    $10^3$  & 1.71 (1.46)  & 0.96 (0.92)  & 1.73 (1.50) \\
    $10^4$  & 1.60 (1.57)  & 0.96 (0.96)  & 1.58 (1.55) \\
    $10^5$  & 1.54 (1.49)  & 0.99 (0.96)  & 1.46 (1.46) \\
 %   $10^6$  & 1.42 (1.36)  & 0.99 (0.95) & 1.34 (1.34) \\
  \hline%  \hline 
      &   \multicolumn{3}{|c|}{{ZS}}    \\ 
    $10^2$  & 1.58 (1.26)  & 1.00 (0.91) & 1.58 (1.38)\\
    $10^3$  & 1.64 (1.39)  & 0.95 (0.92) & 1.73 (1.51)\\
    $10^4$  & 1.55 (1.46)	 & 0.96 (0.95) & 1.61 (1.54)\\
    $10^5$  & 1.37 (1.23)	 & 0.91 (0.83) & 1.51 (1.48) \\
    $10^6$  & 1.10 (0.95)	 & 0.61 (0.55) & 1.80 (1.73)\\
    $10^7$  & 0.89 (0.75)	 & 0.48 (0.43) & 1.85 (1.74)\\ \hline 
%    $10^8$  & 0.79 (0.75)	 & 0.37 (0.33) & 2.14 (1.97)\\  
  \end{tabular}
  \end{table}
%=======================================
One can see that usage of QGSJET-II and SIBYLL models leads to apparent difference of the muon neutrino flux, as well as in the case of SIBYLL as compared to KM  (unlike the muon flux, where SIBYLL and KM lead to very similar results ~\cite{kss08}). On the contrary, the QGSJET-II neutrino flux is very close to the KM one: up to $100$ TeV the difference does not exceed $5\%$ for the GH spectrum and $10\%$  for the  ZS one at $\theta=0^\circ$. 
While the muon flux discrepancy in the QGSJET-II and KM predictions is about $30\%$ at vertical~\cite{kss08}. The origin of differences is evident:  the kaon production ambiguity.  

\begin{figure}[!h]            %
  \centering
\includegraphics[width=70 mm]{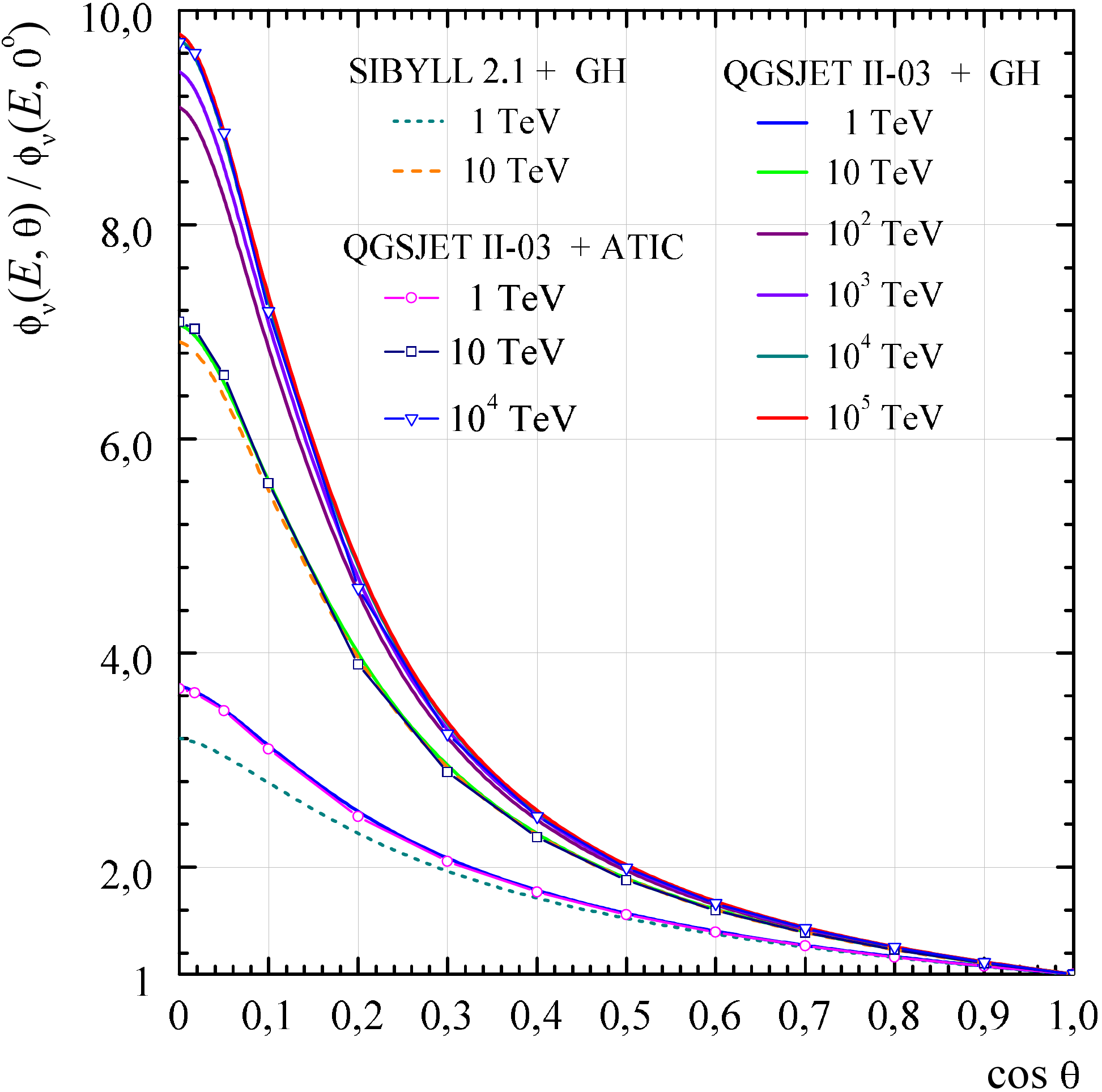} 
 \caption{Zenith-angle enhancement of the $\nu_\mu + \bar{\nu}_\mu$ flux.} 
  \label{angle_distrib} %\vskip -0.5 cm
 \end{figure}

Zenith-angle distributions of the conventional neutrinos, $\phi_{\nu_{\mu}}(E,\theta)/\phi_{\nu_{\mu}}(E,0^\circ)$, for the energy range $1-10^5$ TeV   are shown in Fig.~\ref{angle_distrib}. Calculations are made with  QGSJET{\text -}II and SiBYLL 2.1 models both for GH  and ATIC-2 primary spectra and composition. 
As was expected, a shape of the angle distribution visibly depends on the neutrino energy (at  $E<100$ TeV) particularly  close to horizontal. The effect of hadronic models (as well as of the primary spectrum) on the angle distribution is weak.

%  ================================
%  ================================
 
\begin{figure}[!t]            %
  \centering
\includegraphics[width=70 mm]{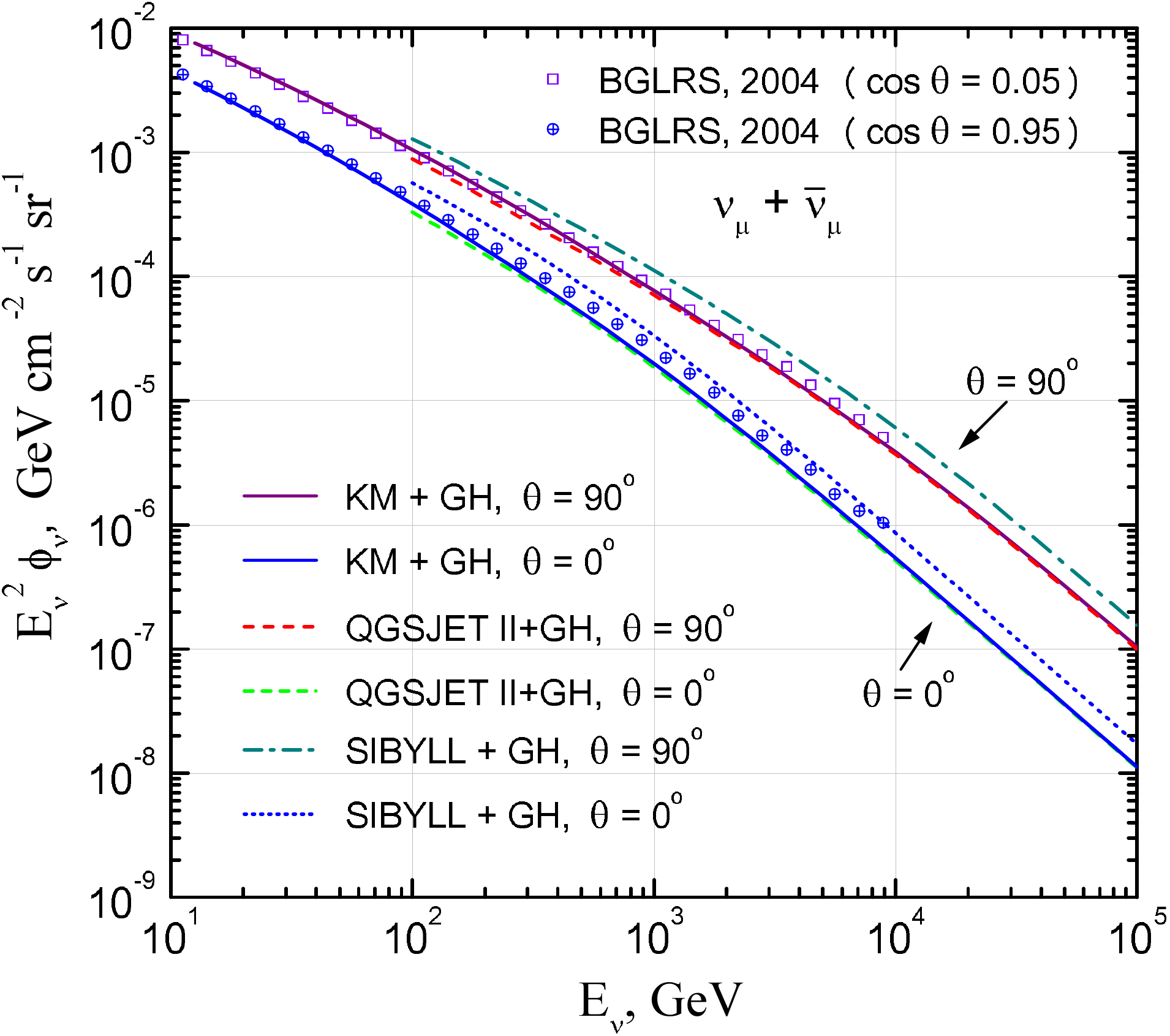}
%  \caption{$(\nu_\mu+\bar\nu_\mu)$ flux calculated with KM, QGSJET-II, SIBYLL 2.1 and TARGET 2.1.}
 \caption{The two independent calculations for the GH spectrum~\protect\cite{GH}.}  \label{3mod_GH} %\vskip -0.5 cm
 \end{figure}

 Figure~\ref{3mod_GH} shows this work calculations of the neutrino flux (lines) in comparison with the result of  Barr, Gaisser, Lipari, Robbins and Stanev (BGLRS)~\cite{BGLRS04}) obtained with use of the  TARGET 2.1 (symbols).  All these computations are performed for the GH primary spectra.  As one can see the calculations for KM and TARGET 2.1 are in close agreement in the range $10-10^4$ GeV (near horizontal) or at $E_\nu < 400$ GeV  near vertical. 
 
\begin{figure}[b]
\centering
\includegraphics[width=70 mm]{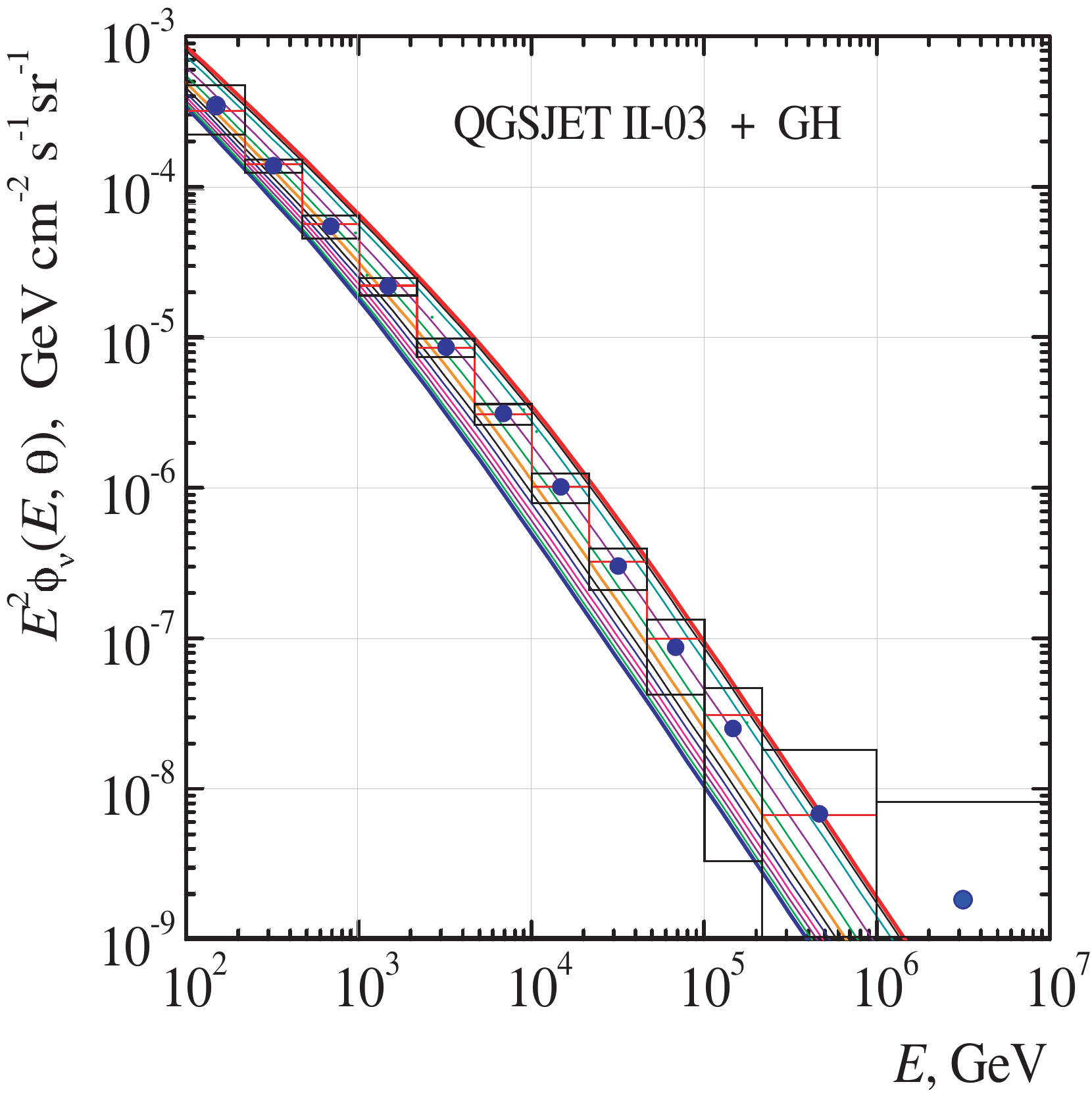}
\caption{Conventional $\nu_\mu+\bar\nu_\mu$  flux at different zenith angles. Blue points: IceCube preliminary muon neutrino spectrum averaged over zenith angles~\cite{icecube09}.}
\label{IceCube_ang}  
\end{figure}
 The calculation of conventional $\nu_\mu+\bar\nu_\mu$  fluxes at different zenith angles
is compared with preliminary data of IceCube experiment in Fig.~\ref{IceCube_ang}.  Curves ($\cos\theta=0\div 1.0$ from top to bottom) display  calculated fluxes  made for GH primary spectra and composition with usage of QGSJET{\text -}II model, blue points with error bars present the IceCube muon neutrino spectrum averaged over zenith angle~\cite{icecube09} (see also~\cite{teresa09}). 
%=======================================
 \begin{figure}[!t]
  \centering \vspace{0.20 cm} % \vskip -1 cm
\includegraphics[ width=70 mm]{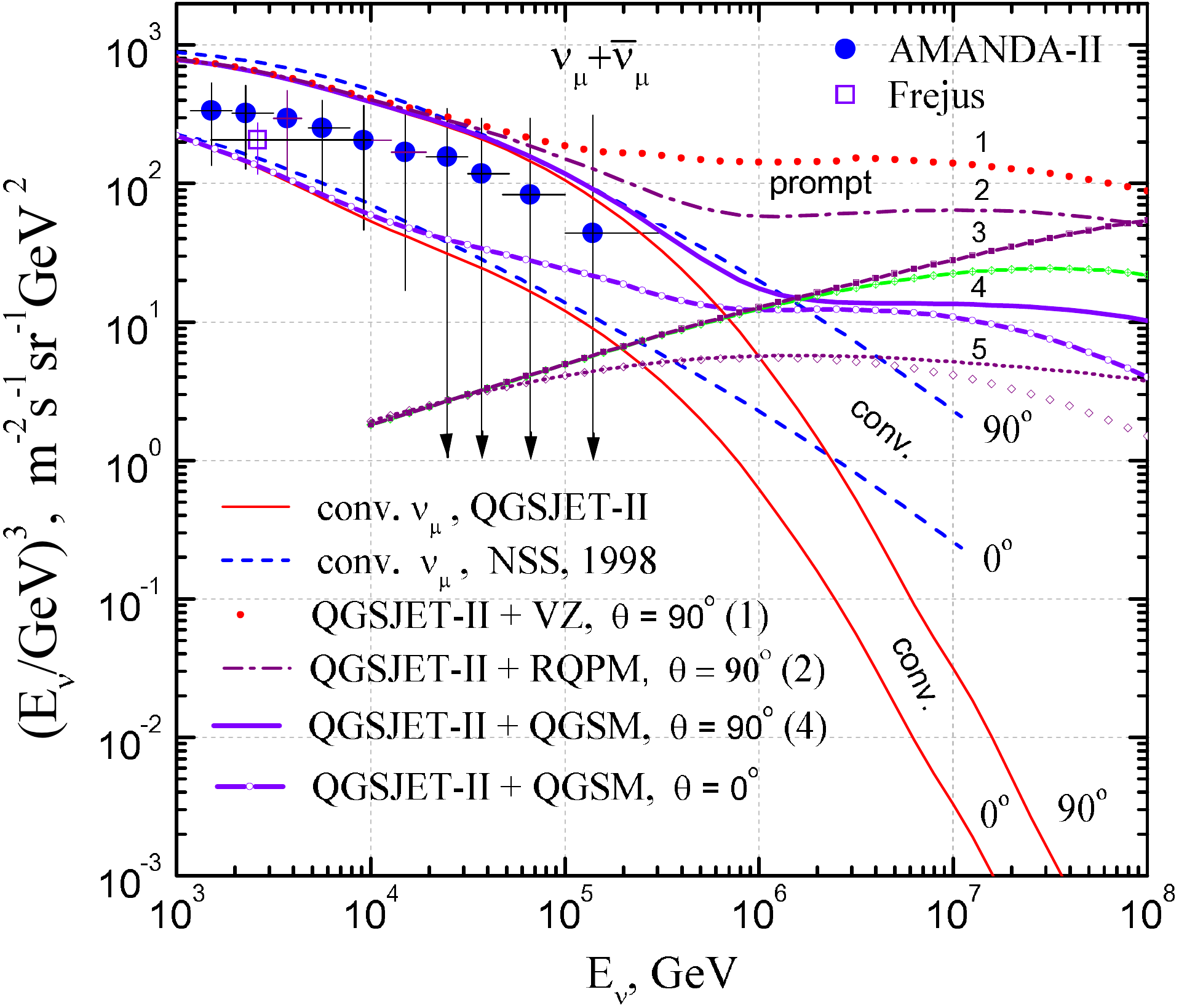} 
 \caption{Fluxes of the conventional and prompt muon neutrinos. Experiments: AMANDA-II~\cite{amanda07} (circles) and Frejus~\cite{frejus} (the square). Calculations: the conventional  flux  --  thin red lines (this work) and dashed~\cite{nss98}; the prompt flux -- VZ~\cite{VZ01} (line 1), RQPM~\cite{bnsz89} (2), GGV ~\cite{GGV00} (3, 5 for $\lambda=0.5, 0.1$), QGSM~\cite{bnsz89} (4).} 
\label{pr_nu_5m} %\vskip -0.2 cm
 \end{figure} 
  
Figure~\ref{pr_nu_5m} presents the comparison of the calculation of the conventional (from $\mu, \,\pi,\, K$-decays) and prompt muon neutrino flux ~\cite{nss98, Naumov2001, bnsz89,  VZ01, GGV00} with the data of the AMANDA-II experiment~\cite{amanda07}. The conventional flux here was computed with use of QGSJET-II model combined with ZS primary spectrum (thin lines ``conv.''). Dashed lines mark the calculation by Naumov, Sinegovskaya and Sinegovsky~\cite{nss98, Naumov2001} of the conventional muon neutrino fluxes for $\theta = 0^\circ$ and $90^\circ$. 
Bold dotted line (1) shows the sum of the prompt neutrino flux by Volkova \& Zatsepin (VZ)~\cite{VZ01} and the conventional one due to the QGSJET-II + ZS model  at $\theta = 90^\circ$.  Dash-dotted line (2) marks the sum of the  QGSJET-II  conventional  flux ($\theta = 90^\circ$) and the prompt neutrino contribution due to the recombination quark-parton model (RQPM)~\cite{bnsz89}. Solid line 4 shows the same for the prompt neutrino flux due to the quark-gluon string model (QGSM)~\cite{bnsz89} (see also~\cite{nss98, Naumov2001, prd98}).  Also shown here  are the two of the prompt neutrino flux predictions by Gelmini, Gondolo and Varieschi (GGV)~\cite{GGV00}: line 3 (5) represents the case of $\lambda=0.5$ ($0.1$), where $\lambda$ is exponent of the gluon distribution at low Bjorken $x$.
Curves just below  3, 4, 5 ones display the coresponding flux at $\theta = 0^\circ$.
\begin{table}[!b]
  \caption{Atmospheric neutrino flux at $E_\nu = 100$ TeV vs. the AMANDA-II restriction for the $\nu_\mu+\bar\nu_\mu$ flux}    
  \label{tab_3}
% \centering % \vspace{0.2cm} 
%\vskip -0.1 cm 
  \begin{tabular}{|l|c|} \hline
     Model & $E_\nu^2\phi_\nu$, (cm$^2$\,s\,sr)$^{-1}$ GeV \\ 
   \hline %\hline 
%    & \multicolumn{3}{c}{GH} \\ 
   conventional $\nu_\mu+\bar\nu_\mu$ : &   $0^\circ$  \quad\quad\quad       $90^\circ$ \\
  QGSJET-II + ZS  & $1.20\times 10^{-8}$ \quad $10.5\times 10^{-8}$ \\   
  QGSJET-II + GH  & $1.11\times 10^{-8}$ \quad $9.89\times 10^{-8}$   \\ \hline  
   prompt $\nu_\mu+\bar\nu_\mu$ : &  $90^\circ$  \\
   QGSM~\cite{bnsz89}         & $1.22 \times 10^{-8}$    \\
   RQPM~\cite{bnsz89}        & $4.61 \times 10^{-8}$  \\
  VZ~\cite{VZ01}           & $8.12 \times 10^{-8}$  \\
  \hline\hline 
 AMANDA-II upper limit~\cite{amanda07}  & $7.4 \times 10^{-8}$   \\
  \hline  
   \end{tabular}  \vskip -0.2 cm
  \end{table}    
Calculated prompt neutrino fluxes  at $E_{\nu}=100$ TeV are presented in Table~\ref{tab_3} along with the upper limit on the  astrophysical muon neutrino diffuse flux obtained in AMANDA-II experiment~\cite{amanda07}. Note that the QGSJET-II+GH flux appears to be the lowest flux of the conventional neutrinos at high energies.   

\section{Summary}

  The calculations of the high-energy atmospheric muon neutrino flux demonstrate rather weak dependence on the primary specrtum models in the energy range $10-10^5$ GeV.  However the picture appears less steady because of sizable flux differences originated from  the models of high-energy hadronic interactions. As it can be seen by the example of the models QGSJET-II and SIBYLL 2.1, the major factor of the discrepancy in the conventional neutrino flux is the kaon production in nucleon-nucleus collisions.  
 
A common hope that atmospheric muon fluxes might be reliable tool to promote the discrimination between the hadronic interaction models seems to be rather illusive as the key differences in the $\pi$/$K$ production impact variously on the neutrino flux and muon one. For the high-energy neutrino production at the atmosphere the kaon yield in nucleon-nucleus interactions is more strong factor in comparison with that for production of the atmospheric muons, despite on their common to neutrinos origin.  

Inasmuch as the atmospheric prompt neutrino flux weakly depends on the zenith angle (near $100$ TeV),  one may refer the AMANDA-II restriction just to the prompt neutrino flux model. Thus one may consider both RQPM and QGSM to be consistent with the AMANDA-II upper limit for diffuse neutrino flux.
  
\begin{acknowledgments}
The work supported by Russian Federation Ministry of Education and Science within the Federal  Programs "Scientific and educational specialists for innovative Russia" under contract numbers P681, P1242 and  "Development of scientific potential in Higher Schools" under grants 2.2.1.1/1483, 2.1.1/1539. 
\end{acknowledgments}

%\bigskip

%\bibliography{Sinegovsky-12_Oct}

 \end{document}